\documentclass{aastex}

\usepackage{natbib}
\usepackage{graphicx,subfigure} % subfigure is needed to put multiple plots in one figure
\usepackage{amsmath}

\usepackage{arydshln}
\usepackage{multirow}

\newcommand\lsim{\la}
\newcommand\gsim{\ga}

\newcommand \Chandra{{\sl Chandra}}
\newcommand \Swift{{\sl Swift}}
\newcommand \newflare{{Swift J174540.7-290015}}
\newcommand \SgrA{Sgr~A*}
\newcommand \GCPWN{PWN~G359.945-0.045}
\newcommand \magnetar{SGR~J1745-29}

\newcommand \mum{ \mu{\rm m} }

\newcommand \NH{{\rm N}_{\rm H}}

\newcommand \col{{\rm cm}^{-2}}

\begin{document}

\title{The Chandra Dust Scattering Halo of Galactic Center transient Swift J174540.7-290015}
\author{L.~R.~Corrales$^{1}$}
\author{B.~Mon$^2$}
\author{D.~Haggard$^2$}
\author{F.~K.~Baganoff$^{3}$}
\author{G.~Garmire$^4$}
\author{N.~Degenaar$^{5,6}$}
\author{M.~Reynolds$^{7}$}
\affil{$^1$Einstein Fellow, University of Wisconsin-Madison, 475 North Charter Street, Madison, WI, 53706, USA}
\affil{$^2$McGill Space Institute, McGill University, 3550 University Street, Montreal, QC, H3A 2A7, Canada}
\affil{$^3$MIT Kavli Institute for Astrophysics and Space Research, 77 Massachusetts Ave, Cambridge, MA, 02139, USA}
\affil{$^4$Huntingdon Institute for X-ray Astronomy, 10677 Franks Road Huntingdon, PA, 16652, USA }
\affil{$^5$Institute of Astronomy, University of Cambridge, Madingley Road, Cambridge CB3 OHA, UK}
\affil{$^6$Anton Pannekoek Institute for Astronomy, University of Amsterdam, Science Park 904, 1098 XH, Amsterdam, the Netherlands}
\affil{$^7$University of Michigan}

\begin{abstract}

We report the detection of a dust scattering halo around a recently discovered X-ray transient, \newflare, which in early February of 2016 underwent one of the brightest outbursts ($F_{\rm X} \approx 5 \times 10^{-10}$~erg~cm$^{-2}$~s$^{-1}$) observed from a compact object in the Galactic Center field.  We analyze four \Chandra\ images that were taken as follow-up observations to \Swift\ discoveries of new Galactic Center transients.  After adjusting our spectral extraction for the effects of detector pileup, we construct a point spread function for each observation and compare it to the GC field before the outburst.  We find residual surface brightness around \newflare, which has a shape and temporal evolution consistent with the behavior expected from X-rays scattered by foreground dust.  
We examine the spectral properties of the source, which shows evidence that the object transitioned from a soft to hard spectral state as it faded below $L_{\rm X} \sim 10^{36}$~erg~s$^{-1}$.  This behavior is consistent with the hypothesis that the object is a low mass X-ray binary in the Galactic Center.  

\end{abstract}

	\section{Introduction}
	\label{sec:Introduction}

The X-ray view of the Galactic Center (GC) is crowded with diffuse emission from hot gas, the supernova remnant Sgr A East, stellar clusters, pulsar wind nebulae, and thousands of X-ray emitting compact objects \citep{Maeda2002, Park2004, Muno2009}.  The only X-ray instrument capable of resolving the most tightly-packed features is the \Chandra\ X-ray observatory, which has $0.5''$ per pixel resolution and a point spread function (PSF) that confines about 90\% of source light to a $2''$ region.  
However, with hydrogen column density of $\NH \sim 10^{23}\col$, the sight line towards the GC is optically thick to both  absorption and scattering of soft X-rays, with $\tau_{\rm sca}(1~{\rm keV}) \sim 5$.  Due to absorption, X-ray images of the GC rarely show any signal below 2~keV.  
Because dust also bends X-ray light through arc-minute scale angles, the scattered light is often recaptured by a telescope, appearing as a diffuse dust scattering `halo' image \citep[e.g.][]{Rolf1983,PS1995,VS2015}.  For the 2-6~keV range ($\tau_{\rm sca} \sim 0.1 - 1$), our image of the GC is effectively blurred by stray dust-scattered light.  

Understanding the X-ray obscuration properties of the interstellar medium (ISM) is crucial for modeling accretion from compact objects as well as for learning about the material properties of dust \citep{Smith2016, Corrales2016}.  In GC studies, the ISM column obtained from fitting the X-ray continuum spectrum is often used to make conclusions about where the compact object is in relation to \SgrA\ \citep[e.g.][]{Kennea2013}.  Furthermore, the ISM column obtained from X-ray absorption does not match with that obtained from extinction measurements in the infrared \citep{Fritz2011}; they differ by a factor of two.  
Finally, \Chandra\ observations have revealed that the X-ray emission from \SgrA\ is extended, allowing us to test theoretical models that trace the flow of hot gas as it falls into the supermassive black hole \citep{Baganoff2003, Shcherbakov2010, Wang2013, Roberts2016}.  This situation is complicated by the fact that \SgrA\ is variable, flaring on the order of once per day for hours at a time \citep[e.g.][]{Neilsen2013} and impacting neutral gas in the vicinity of the GC \citep[e.g.][]{Krivonos2016}.  The dust scattering halo around \SgrA\ will also vary temporally based on the light curve of \SgrA\ and the locations of foreground dusty material \citep[e.g.][]{TS1973,Tiengo2010,Heinz2015,Heinz2016}.

We seek to characterize the dust in the foreground of the GC, which is especially important for examining the extended image of \SgrA.  
In addition to \SgrA, there are may other accreting compact objects that often brighten dramatically in the X-ray, providing many opportunities to observe dust scattering from the GC foreground.  Since February 2006, the \Swift\ X-ray telescope has been monitoring the GC with 1~ks observations every 1-4~days \citep[e.g.][for a recent review]{Degenaar2015}.  
Over the years, this monitoring campaign has detected the magnetar \magnetar, several X-ray flares from \SgrA, and numerous outbursts of transient X-ray sources, of which 5 previously unknown systems. Two of those previously unknown X-ray transients were discovered in 2016: \newflare\ \citep{ATEL8649} and Swift J174540.2-290037 \citep{ATEL9109}.

We report here on the detection of dust scattering around \newflare, a compact object $16''$ North-East of \SgrA, which underwent the 
second brightest outburst observed in the GC field \citep[Table 3 of][]{Degenaar2015}.  
Analysis of the \Swift, XMM-Newton, and INTEGRAL observations during the first part of the outburst (February and March of 2016) show that the transient underwent significant changes in its spectral shape, softening from a photon index of $\Gamma \sim 2$ to $\Gamma \sim 6$ \citep[][hereafter P16]{Ponti2016}.  
Such a spectral evolution is very common for the outbursts of transient low mass X-ray binaries (LMXBs), which initially start in a``hard spectral state'' that is dominated by the emission from a hot electron plasma (referred to as a corona), but move to a ``soft spectral state'' around the peak of the outburst when thermal (i.e. soft) emission from the accretion disk dominates the X-ray spectrum.  During the decay from the outburst peak, these objects transition back to a hard spectral state before returning to quiescence.  Based on their spectral analysis, P16 proposed that \newflare\ is a LMXB.  

In this work, we describe the \Chandra\ observations made in response to the \Swift\ discoveries of two transients in 2016  and the data analysis in Section~\ref{sec:Data}.  We find residual surface brightness around \newflare, and perform several tests to confirm that it is dust scattering in Section~\ref{sec:ScatteringHalo}.  We fit the \Chandra\ spectra of \newflare\ and compare our results to the work of P16 in Section~\ref{sec:Spectrum}.  We summarize our conclusions and mark out future work in Section~\ref{sec:Conclusions}.

	\section{Data Reduction and Analysis}
	\label{sec:Data}

\begin{table}
\caption{\Chandra\ observations used in this work}
\label{tab:ObsIDs}
\centering
\begin{tabular}{l c c c l}
	\hline
	{\bf ObsID}	& {\bf Exposure}		& {\bf Start Date} 	&	{\bf (MJD)}		& {\bf Instrument} \\
	\hline
	3392			& 166.7~ks			& 2002-05-25		&	(52419)	& ACIS-I \\	
	16216			& 42.7~ks			& 2014-08-02		&	(56871)	& ACIS-S (1/8) \\	
	\hline	
	18055		& 22.7~ks			& 2016-02-13		&	(57431)	& ACIS-S (1/8) \\
	18056		& 21.8~ks			& 2016-02-14		&	(57432)	& ACIS-S (1/8) \\
	18731			& 78.4~ks			& 2016-07-12		&	(57582)	& ACIS-S (1/8) \\
	18732		& 76.6~ks			& 2016-07-18		&	(57587)	& ACIS-S (1/8) \\
	\hline
\end{tabular}
\end{table}

In February 2016, the \Swift\ X-ray observatory detected an outburst from a new object, \newflare, about $16''$ north-east of \SgrA.  Assuming a distance of 8~kpc, the initial outburst reached a 2-10~keV luminosity around $8 \times 10^{35}$~erg/s \citep{ATEL8649}.  The source continued to brighten, and was observed about a week later with the \Chandra\ X-ray telescope in ACIS-S, 1/8~subarray mode (ObsIDs 18055 and 18056).  In July 2016, the discovery of Swift J174540.2-290037 \citep{ATEL9109} triggered a \Chandra\ observation of the GC field (ObsIDs 18731 and 18732), also in 1/8~subarray mode.  \newflare\ was still bright around this time, offering further coverage of its spectrum and dust scattering halo, five months after the initial detection.  For all of the data analysis described below, we used CIAO version 4.8 with CALDB version 4.7 for image reduction and the {\sl Interactive Spectral Interpretation System} \citep[ISIS;][]{ISISsoftware} version 1.5 for spectral fitting.

	\subsection{Surface Brightness Profiles}
	\label{sec:Profiles}

\begin{figure}
	\includegraphics [width=\columnwidth, trim=0 0 0 0] {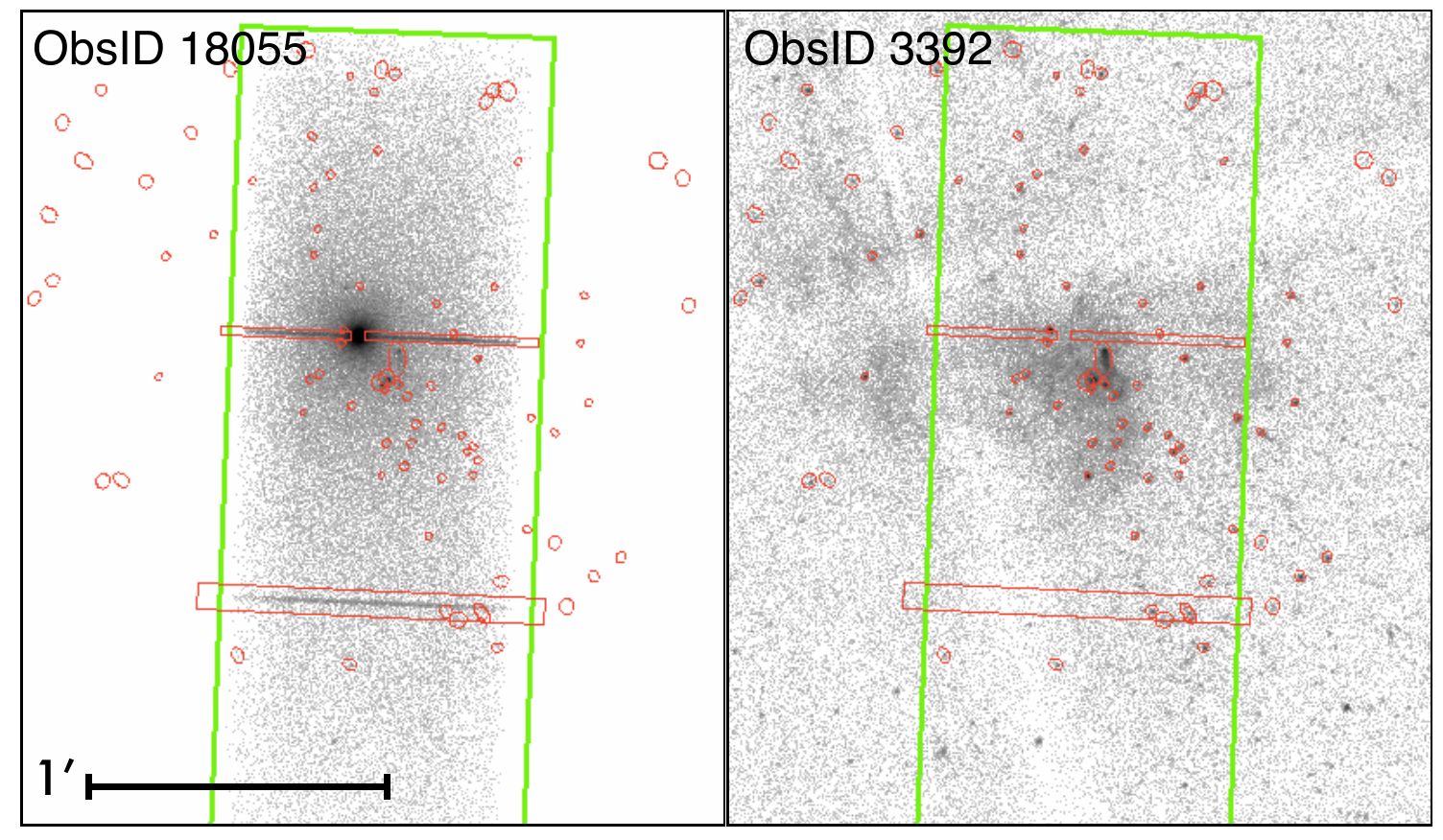}
	\caption{A \Chandra\ ACIS-S image of \newflare\ (left) compared to a deep ACIS-I observation (right), taken before the outburst. For each outburst image, we removed readout streaks and sources identified from ObsID 3392 (red regions) before extracting a surface brightness profile.  We excluded identical regions from ObsID 3392 and only included events within the field of view of the ACIS-S subarray (green region).  This process was repeated for each ACIS-S observation of \newflare, producing four pairs of surface brightness profiles, before and after the outburst (Figure~\ref{fig:RadialProfiles}).}
	\label{fig:Images}
\end{figure}

\begin{figure*}[tp]
	\centering
	\includegraphics [width=\textwidth, trim=0 0 0 0] {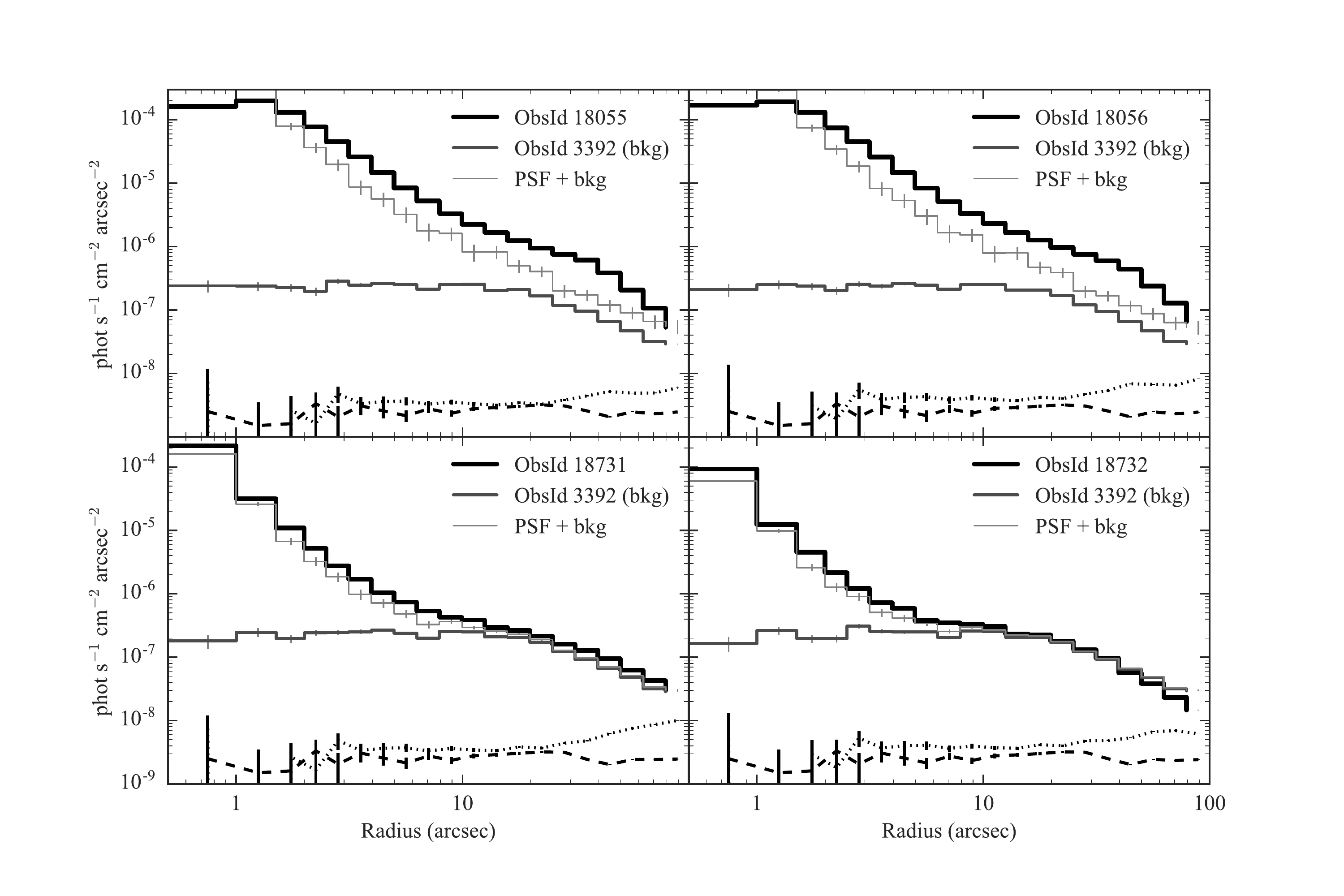}
	\caption{
	The 1-6~keV intensity profiles for \newflare\ (thick black), the respective background profile from ObsID 3392 (thick grey), and the background plus dust-free PSF (thin grey), constructed using the template method of \citet{CorralesCygX3}.  The contributions from the charged particle background are shown for ACIS-I (black dashed) and ACIS-S (black dotted).
	}
	\label{fig:RadialProfiles}
\end{figure*}

For each \Chandra\ observation of \newflare\ (Table~\ref{tab:ObsIDs}), we applied the following methods to get a surface brightness profile.  We centered an annular region file on the mean pixel coordinates from the events within a $10''$ radius circle of \newflare.  The annular bins are 1~pixel wide in the inner portion and log-spaced from 6-200 pixels, to match the PSF templates in \citet{CorralesCygX3}.  Before extracting the profiles, we removed readout streaks, other bright objects, and all point sources in the field (Figure~\ref{fig:Images}, described in more detail below).  We used the  CIAO \texttt{fluximage} script to create 2.75~keV exposure maps, the approximate energy at which the dust scattering halo is expected to peak.  Then we extracted 1-6~keV profiles, applying the exposure map with the CIAO \texttt{dmextract} tool (Figure~\ref{fig:RadialProfiles}, thick black lines).

We selected a deep (160 ks) ACIS-I image of the GC (ObsID 3392) for comparison of the GC field radial profile before and after a transient outburst.  We used the CIAO tool \texttt{celldetect} to identify point sources in ObsID 3392, which can also be variable.  We used the resulting regions to remove all nearby point sources.  In addition, we created custom region files for each observation in order to remove readout transfer streaks, \SgrA, the nearby magnetar \magnetar, and \GCPWN.  
We removed identical readout streak regions from ObsID 3392 before extracting the radial profiles from the location of \newflare.  We also applied a window to the ObsID 3392 field of view so that it corresponded to the field of view covered by each of the 1/8 subarray ACIS-S observations.  
The results can be seen in Figure~\ref{fig:RadialProfiles} (thick grey lines).

We next adjusted our profiles for the charged particle background, which accounts for 1-10\% of the 1-6~keV surface brightness within $100''$ of \SgrA.  In addition, the front illuminated ACIS-I and back illuminated ACIS-S chips are affected differently by charged particles.  We extracted 1-6~keV radial profiles (with no exposure map) from the stowed ACIS-I and ACIS-S background files, reprojected for a GC pointing.  The charged particle flux can also vary slowly over time due to Solar activity, so we scaled the stowed background to match the 10-14~keV background in each observation of \newflare.  Figure~\ref{fig:RadialProfiles} shows the ACIS-I ( black dashed line) and ACIS-S (black dotted line) contributions to the surface brightness profiles.  The charged particle background accounts for 10-30\% of the flux on the edges of our profile.

	\subsection{Source spectrum and pileup corrections}
	\label{sec:SourceSpectrum}

The \Chandra\ ACIS instrument contains an array of CCDs with no shutter.  For bright sources, many X-ray photons can deposit charge in the same region of the detector before the CCD is read out.  This leads to the phenomenon of pileup, for which multiple photons are mistaken for a single photon of higher energy.  With pileup, one cannot assume a direct relation between the CCD electron count and photon energies, making it difficult to characterize the source spectrum.  Piled spectra typically exhibit an excess of high energy photons.  In addition, pileup can cause an overall reduction in the apparent flux, because piled events may be confused with a cosmic ray event, which are rejected by the on-board system or in later calibration pipelines.  

\begin{figure*}[tp]
	\centering
	\includegraphics [width=\textwidth, trim=0 0 0 0] {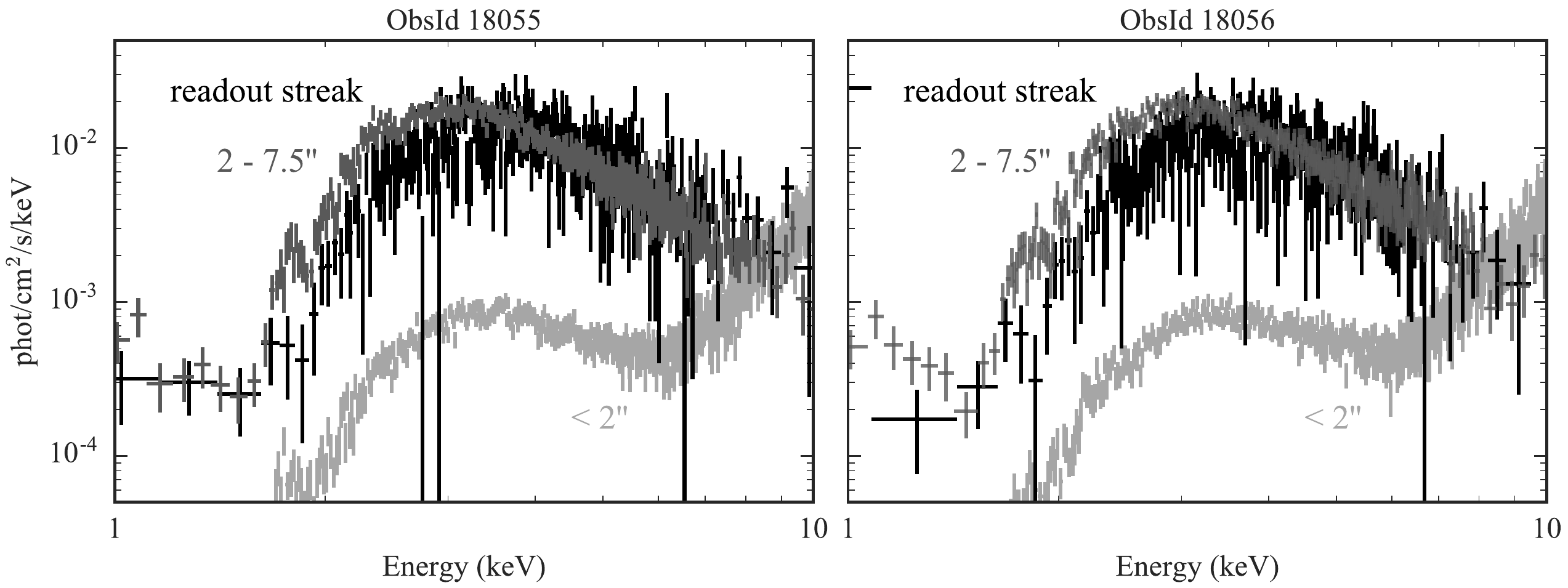}
	\caption{
	Flux spectrum of \newflare\ extracted using three methods.  
	{\sl Light grey} shows the CIAO \texttt{specextract} results within a circular aperture of $2''$ radius.
	{\sl Medium grey} shows the modified \texttt{specextract} from an annulus spanning 2-7.5$''$, which is relatively pileup free, but affected by dust scattering.
	{\sl Black} shows the readout streak spectrum.  We expect the readout streak spectrum to be the least affected by pileup and dust scattering.
    	}
    	\label{fig:SourceSpectrum}
\end{figure*}

We extracted a spectrum of \newflare\ from the central region of the source ($r < 2''$) using CIAO \texttt{specextract}.  
To characterize the background, we used a fairly long (43 ks) ACIS-S image of the GC (ObsID 16216), in which the transient was not active.  
We removed the regions containing known sources, as described in Section~\ref{sec:Profiles}, before extracting spectra from the position of \newflare.  
The high energy tail apparent in the $r < 2''$ spectrum is a sure signature of pileup (light grey in Figure~\ref{fig:SourceSpectrum}), so we used other methods to get a reliable spectrum.  
For the brightest sources, a spectrum can be extracted from the readout streak.  We followed the CIAO science thread,\footnote{http://cxc.harvard.edu/ciao/threads/streakextract/} and chose regions adjacent on each side of the readout streaks to describe the background.  The results are shown in Figure~\ref{fig:SourceSpectrum} (black).

For sources that are not bright enough to produce a bright readout streak, one can extract a spectrum from an annulus around the point source, which captures the PSF wings.  In order to choose a region around \newflare\ that was not affected by pileup, we examined the 10-14~keV image, which contains only uniformly distributed background and piled events from the center of the brightest sources.  We chose an annular region spanning $2'' - 7.5''$, which covers an appreciable amount of the PSF wings without overlapping other point sources.  However, we found that the default auxiliary response function (ARF) produced by \texttt{specextract} for this region needed to be corrected.  We assigned the ARF obtained by subtracting the $r < 2''$ psf-weighted ARF from the $r < 7.5''$ psf-weighted ARF for our source.  We checked the validity of this method by measuring the fraction of counts captured within the same annulus in a MARX simulated PSF.  The two methods for calculating the enclosed PSF fraction agreed on the 1\% level.

Figure~\ref{fig:SourceSpectrum} (medium grey) shows that the annular spectrum, with the ARF correction described above, is in good agreement with the readout streak spectrum on the high energy side.  The excess of soft energy photons occurs because the annular spectrum is capturing portions of the dust scattering halo, which is brighter at softer energies.  Because the annular spectrum is contaminated with dust scattering, we consider the readout streak method to be more reliable.  
To describe the apparent flux of \newflare, we use the readout streak for ObsIDs 18055 and 18056.  We use the annular spectrum for ObsIDs 18731 and 18732, because \newflare\ was too dim to produce a bright readout streak in those two observations.

We fit the continuum spectrum of \newflare\ with a power law attenuated by both ISM absorption \citep[\texttt{tbnew}, ][]{Wilms2000} and dust scattering, where $\tau_{\rm sca} = 0.5 (\NH/10^{22}\col)$.\footnote{The choice for this scaling relation is described in more detail in Section~\ref{sec:Spectrum}.}  This smoothed model spectrum was used to estimate the flux for PSF construction (described below).  The $\NH$ column obtained from the power law fits to the readout streaks are consistent with other GC sources, including \SgrA, strongly suggesting that \newflare\ is part of the population of GC compact objects \citep[e.g.][]{Baganoff2003, Nowak2012}.  
Table~\ref{tab:Flux} lists the absorbed and unabsorbed fluxes, along with the source luminosity assuming it is in the GC.

\begin{table*}
\centering
\caption{Flux results for Swift J174540.7-290015, fit with power law and ISM extinction}
\label{tab:Flux}
\begin{tabular}{l c c c l}
	\hline
	{\bf ObsId}	& {\bf Absorbed Flux$^a$} 	& {\bf Unabsorbed Flux$^a$}	& {\bf Luminosity$^b$ (8 kpc)} 	&  {\bf method}  \\
	\hline
	18055	& 3.43 (3.32, 3.55) $\times 10^{-10}$		& 3.76 (3.36, 4.24) $\times 10^{-9}$		& 2.89 (2.58, 3.25) $\times 10^{37}$	& readout \\
	18056	& 3.15 (3.04, 3.26) $\times 10^{-10}$		& 3.67 (3.27, 4.15) $\times 10^{-9}$		& 2.82 (2.51, 3.19) $\times 10^{37}$ & readout \\
	18731	& 2.13 (2.16, 2.10) $\times 10^{-11}$		& 1.13 (1.06, 1.21) $\times 10^{-10}$	  	& 8.68 (8.14, 9.31) $\times 10^{35}$	& annulus	\\
	18732	& 6.74 (7.31, 6.22) $\times 10^{-12}$		& 4.55 (3.92, 5.44) $\times 10^{-11}$ 	& 3.49 (3.01, 4.17) $\times 10^{35}$	& annulus	\\
	\hline
		& \multicolumn{2}{l}{$^a$erg/s/cm$^2$ (2-10 keV)}	& $^b$erg/s (2-10 keV)	& \\
	\hline
\end{tabular}
\end{table*}

	\subsection{PSF construction}
	\label{sec:PSF}

With the apparent flux spectrum of \newflare, we constructed the PSF profile using the template method described in \citet{CorralesCygX3}.  This method is preferred for \Chandra\ ACIS-S targets because it has been shown that the ray-tracing software is inaccurate for the \Chandra\ PSF wings \citep{Smith2008}.  Integrating the spectrum over 0.5~keV wide bins between 1 and 6~keV, we used the flux-normalized intensity profile from a \Chandra\ HETG image of QSO B1028+511 (ObsID 3472) to construct a dust-free PSF for \newflare.  This observation was chosen because it contains a bright X-ray point source with a very low dust column ($\NH < 10^{20}$~cm$^{-2}$) and exhibits the least amount of pileup ($\sim 5\%$) of the three PSF template candidates in \citet{CorralesCygX3}.

Figure~\ref{fig:RadialProfiles} shows the radial intensity profiles for \newflare\ in each observation (thick black) alongside the respective PSF reconstruction (thin grey) and background estimation from ObsID 3392 (thick grey).  The central portion of the reconstructed PSF is brighter than \newflare\ in Obsids 18055 and 18056 because of pileup.  There is a clear indication of residual brightness in the outburst observations, which fade to match the background as \newflare\ also fades.  The most likely explanation is X-ray scattering from dust in the GC foreground, due to intermediate ISM along the sight line.

	\section{Confirmation of a dust scattering halo}
	\label{sec:ScatteringHalo}

\begin{figure}[tp]
	\centering
	\includegraphics [width=0.48\textwidth, trim=0 0 0 0] {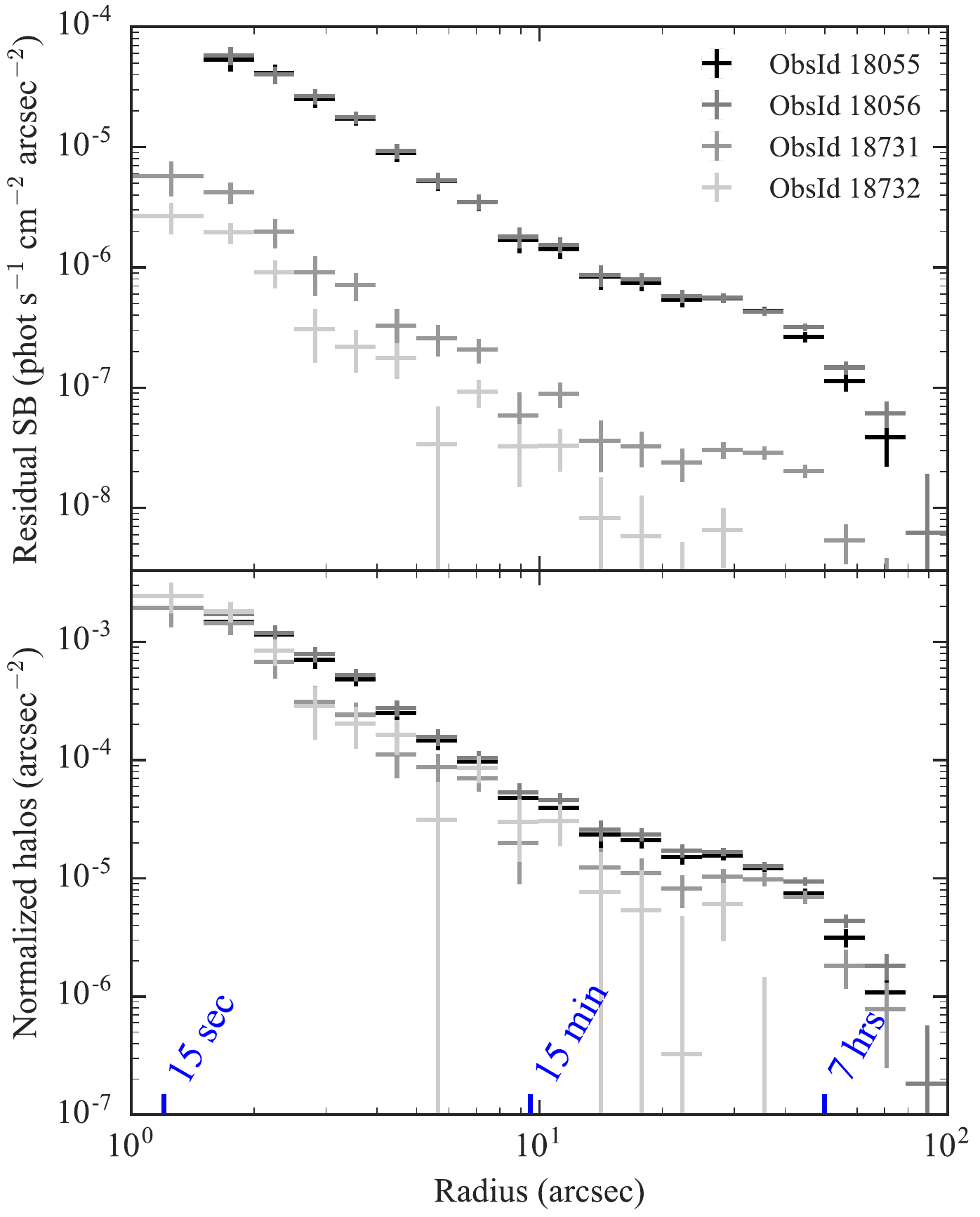}
	\caption{
	{\sl Top:} Residual intensity profile for the dust scattering halos.
	{\sl Bottom:} Normalized profile for the dust scattering halos.  
	Time delays for dust situated half way along the sight line ($x=1/2$) are marked along the bottom of the plot.
	}
	\label{fig:ScatteringHalo}
\end{figure}

Figure~\ref{fig:ScatteringHalo} (top) shows the residual intensity profile after subtracting the template PSF and background.  Negative residuals, which are caused either by pileup or by negligible scattering halo surface brightnesses, are not included.  The spectrum from the first two observations is remarkably stable -- in a 0.5~keV wide bin-by-bin comparison, the model spectrum (Table~\ref{tab:Flux}) of \newflare\ varies by $<10\%$ between ObsiD 18055 and 18056 -- so the residual surface brightness was also very steady in the first two observations.  In the later observations, one can see that the residuals decline as the point source also declines in brightness.

For steady sources that do not vary with time, the dust scattering halo intensity will be directly proportional to the point source's apparent flux ($F_a)$ by the integral:
\begin{equation}
\label{eq:Ihalo}
	I_h(\alpha, E) = F_a(E) \int \int \frac{d\sigma}{d\Omega}(\alpha, E, a, x) \frac{dN}{da}(x)~da~dx
\end{equation}
where $\alpha$ is the observed angular distance between the point source and a scattered photon,  $dN/da$ describes the dust grain size ($a$) distribution, $E$ is the photon energy, and $x \equiv d/D$ describes the position of the dust grains along the sight line, where $D$ is the distance to the point source and $d$ is the distance to the scatterer \citep[e.g.][]{MG1986,SD1998,CorralesCygX3}.  

Figure~\ref{fig:ScatteringHalo} (bottom) shows the normalized intensity profile ($I_h/F_a$) for each observation.  The magnitude and shape of all four normalized profiles agree relatively well, 
as expected for a dust scattering halo.  
The morphology of the surface brightness profiles suggests that there is a wall of dust responsible for the scattering halo's relative flatness around 10-25$''$.  The more highly sloped portion of the profile ($<10''$) is likely to come from dust that is very close to the GC or from a relatively uniform distribution of dust.  For a full discussion on the morphology of dust scattering halos based on the line of sight dust, see \citet{CorralesCygX3} and \citet{VS2015}.  

We attempted to fit the dust scattering halo of ObsID 18055 with a static, optically thin model utilizing the RG-Drude approximation \citep{MG1986,SD1998}.  We found that the $\alpha <10''$ surface brightness could be explained by a power law distribution of dust grain sizes consistent with that describing Milky Way extinction properties \citep[][hereafter MRN]{MRN1977}.  Even though the estimated ISM column towards GC objects is $\NH \approx 1-1.5 \times 10^{23}~\col$, only a small amount of dust is needed explain this inner scattering halo: an ISM column of about $\NH = 4.4 \times 10^{21}~\col$ \citep[assuming a dust-to-gas mass ratio of 1/100;][]{DraineBook} located within 250~pc of the GC.  This represents only a few percent of the total ISM along the sight line.  
For the outer portion of the scattering halo ($\alpha > 10''$), we were unable to find solutions that did not require a highly unusual distribution of dust grains ($a > 0.5~\mum$) to explain the rapid fall in surface brightness at the edge of the profile.  Dust grains of this size break many of the approximations used to speed the computation of X-ray scattering halos.  
Our current model also does not address non-uniformities in the azimuthal dust distribution that can lead to enhanced (due to scattering) or diminished (due to absorption) changes in surface brightness \citep[e.g.][]{Heinz2015}.  
More importantly, the scattering halo surface brightness profile is sensitive to source variability.

Note that Equation~\ref{eq:Ihalo} is technically only valid for steady (non-variable) sources.  Due to traveling a longer distance, dust-scattered photons received by the telescope are delayed relative to the source photons by a time:
\begin{equation}
\label{eq:TimeDelay}
	t' = \alpha^2~\frac{D}{2c} \frac{x}{(1-x)}.
\end{equation}
For X-ray sources in the GC region, approximately 8~kpc away \citep[e.g.][]{Boehle2016}, we can expect
\begin{equation}
\label{eq:TimeDelayUnits}
	t' \approx 10 \left(\frac{\alpha}{{\rm arcsec}}\right)^2 \frac{x}{(1-x)}~{\rm seconds}
\end{equation}
based on small angle geometry.  Some benchmark time delays for ISM at the intermediate value of $x = 1/2$ are overlaid on Figure~\ref{fig:ScatteringHalo}.  For a fixed observation angle, scattering from dust located closer to the observer will incur a shorter time delay, and scattering from dust closer to the GC will incur a longer time delay.

One can expect time variations by the point source to produce deviations from the steady state scattering halo profile.  As discussed above, the spectrum of \newflare\ was remarkably stable between ObsID 18055 and 18056.  That stability is also evident in the scattering halo residuals, which differ by less than one sigma.  
The later two observations (ObsID 18731 and 18732) follow a similar shape but deviate from the earlier observations in the 2-5$''$ and 10-30$''$ range.  This is to be expected if the object underwent any variations over the time scale of hours.  

We extracted light curves from the readout streak (ObsIds 18055 and 18056) and the annular regions used to extract spectra (ObsIds 18731 and 18732), in order to search for obvious signs of variability that could explain the scattering halo differences.  There are no clear sinusoidal trends or strong flares in any of the light curves, so we fit a linear model to each.  For ObsIds 18055 and 18056, our linear fits are consistent with a slope of zero, and the y-intercepts are identical to within $1\sigma$.  For ObsIds 18731 and 18732, our best fit linear model yields a 25\% and 35\% decay in the source brightness over the course of each observation ($\sim 20$~hours).  For all light curves examined, there is a standard deviation of about 10-15\% from the best fit model.  The long term evolution of ObsIds 18731 and 18732 might explain the evolution of the scattering halo at larger scattering angles.  

We plan a follow up study to compare the scattering halo and source light curve, which will place a much stronger constraint on the line of sight position of dust clouds.  For example, if the inner scattering halo really can be explained by dust within 250~pc of the GC, we expect a time delay of about 2 hours.  For the outer edge of the halo, we expected time delays $\gsim 7$~hours.

	\section{Fits to the \newflare\ Spectrum}
	\label{sec:Spectrum}

\begin{table}
\centering
\caption{Fits to \newflare\ spectra from \Chandra}
\label{tab:AllFits}
\begin{tabular}{l c c c c c}
	\hline
	{\bf }	& {\bf Fit}	& ${\rm N}_{\rm H}$		& $\Gamma$	& $kT$		&  $\chi^2/\nu$	\\
		&		& {\footnotesize $10^{22}$~cm$^{-2}$}	&			& {\footnotesize keV}		&  \\
	\hline
	\multicolumn{6}{c}{\bf ObsId 18055} \\
	
	\multirow{3}{*}{\sl annulus} 

 & 1 & $13.1_{-0.2}^{+0.2}$ & $4.55_{-0.07}^{+0.08}$ & ... & 466.1/373\\

 & 2 & $7.8_{-0.1}^{+0.1}$ & ... & $0.75_{-0.01}^{+0.01}$ & 843.1/373\\

 & 3 & $12.6_{-0.3}^{+0.6}$ & $3.0_{-0.3}^{+0.3}$ & $0.40_{-0.03}^{+0.02}$ & 356.4/371\\

\hdashline[0.5pt/5pt]

	\multirow{3}{*}{\sl readout}
 & 1 & $14.9_{-0.8}^{+0.9}$ & $3.9_{-0.2}^{+0.2}$ & ... & 291.2/276\\

 & 2 & $9.2_{-0.6}^{+0.5}$ & ... & $0.91_{-0.04}^{+0.04}$ & 312.7/276\\

 & 3 & $11.6_{-0.4}^{+1.0}$ & $-0.03_{-0.45}^{+1.34}$ & $0.68_{-0.06}^{+0.05}$ & 280.7/274\\
	
	\hline
	\multicolumn{6}{c}{\bf ObsId 18056} \\

	\multirow{3}{*}{\sl annulus}
	
 & 1 & $13.0_{-0.2}^{+0.2}$ & $4.50_{-0.08}^{+0.08}$ & ... & 537.6/377\\

 & 2 & $7.6_{-0.1}^{+0.2}$ & ... & $0.76_{-0.01}^{+0.01}$ & 877.7/377\\

 & 3 & $12.9_{-0.3}^{+0.4}$ & $3.2_{-0.2}^{+0.2}$ & $0.39_{-0.01}^{+0.02}$ & 440.6/375\\

\hdashline[0.5pt/5pt]

	\multirow{3}{*}{\sl readout}

 & 1 & $15.2_{-0.9}^{+0.9}$ & $4.0_{-0.2}^{+0.2}$ & ... & 242.6/263 \\

 & 2 & $9.2_{-0.6}^{+0.5}$ & ... & $0.89_{-0.03}^{+0.05}$ & 260.4/263\\

 & 3 & $14.6_{-1.3}^{+4.9}$ & $3.2_{-1.3}^{+2.7}$ & $0.46_{-0.06}^{+0.06}$ & 240.4/261\\

	\hline
	\multicolumn{6}{c}{\bf ObsId 18731} \\
	
	\multirow{3}{*}{\sl annulus} 
 
 & 1 & $8.5_{-0.4}^{+0.4}$ & $2.5_{-0.1}^{+0.1}$ & ... & 316.6/310\\

 & 2 & $4.9_{-0.3}^{+0.2}$ & ... & $1.19_{-0.04}^{+0.03}$ & 380.7/310\\

 & 3 & $10.6_{-0.7}^{+1.4}$ & $1.8_{-0.2}^{+0.3}$ & $0.40_{-0.06}^{+0.05}$ & 288.4/308\\

%\hdashline[0.5pt/5pt]
	
	\hline
	\multicolumn{6}{c}{\bf ObsId 18732} \\
	
	\multirow{3}{*}{\sl annulus}

 & 1 & $9.6_{-1.1}^{+1.3}$ & $2.2_{-0.3}^{+0.2}$ & ... & 218.0/211\\

 & 2 & $5.4_{-0.7}^{+0.8}$ & ... & $1.42_{-0.08}^{+0.09}$ & 229.1/211\\

 & 3 & $17.4_{-2.8}^{+0.0}$ & $1.8_{-0.6}^{+0.4}$ & $0.35_{-0.08}^{+0.10}$ & 208.2/209\\

%\hdashline[0.5pt/5pt]
	
	\hline
\end{tabular}

\tablecomments{{\footnotesize Model 1 is a power-law; Model 2 is a single temperature blackbody; Model 3 is a power-law plus single temperature black body.  All fits include extinction from ISM absorption (tbnew) and dust scattering (dustscat).}}\end{table}

For completeness, we fit the low resolution \newflare\ continuum spectrum with three basic models: a single power law, a single black body, and a power law plus black body.  All spectra were grouped to obtain minimum of 10 counts per bin.  All fits included the effect of ISM absorption using  \texttt{tbnew}\footnote{http://pulsar.sternwarte.uni-erlangen.de/wilms/research/tbabs/} and the effects of dust scattering removing light from the source aperture.  Because the standard scattering model \texttt{dust}\footnote{https://heasarc.gsfc.nasa.gov/xanadu/xspec/manual/node227.html}, distributed by XSPEC, only works for the optically thin case, we use the custom dust scattering model from \citet{Baganoff2003}, which applies an $\exp({-\tau})$ extinction term.  In all cases, the 1~keV optical depth of dust scattering was tied to the $\NH$ value assuming $\tau_{\rm sca} = 0.5 (\NH/10^{22}~{\rm cm}^{-2})$.  This value is slightly more than that used in \citet{Nowak2012}, who adjusted the dust scattering optical depth of \citet{PS1995} to account for the ISM metal abundances of \citet{Wilms2000}.  We use the optical depth as calculated in \citet{Corrales2016}, which is also consistent with the ISM metal abundances of \citet{Wilms2000}, but the scattering cross-section value is the theoretical one implied by the MRN grain size distribution.  

The RG-Drude ($E^{-2}$ dependence) approximation is known to overestimate the true dust scattering cross section at low energies, particularly $\lsim 1$~keV \citep{SD1998,CorralesCygX3,Corrales2016}.  However, because the ISM column is so large, GC point sources are generally not visible at energies $< 1.5$~keV (Figure~\ref{fig:SourceSpectrum}).  Hence, the signal we are working with to fit the continuum spectrum is well served by the RG-Drude approximation.

The dust scattering model for continuum fitting also attempts to correct for the fact that some dust scattered light will be re-captured by the source extraction aperture.  It assumes that all the dust scattered light is distributed evenly within a circular disk with radius proportional to $E^{-1}$.  
With the default parameter values, less than 5\% of the dust scattering halo is included in the source spectrum, affecting the spectrum overall by less than 1\%.  
However, it has been shown that the disk-shaped approximation vastly underestimates the true amount of the scattering halo recaptured within a typical source extraction radius; it is also heavily affected by the dust grain size and spatial distribution \citep{Smith2016,Corrales2016}.  
Performing the full dust scattering calculations is time consuming and provides many degenerate solutions.  We defer a full continuum fit for dust scattering until after we discover the locations of the dust scattering clouds, breaking the degeneracy between grain size and location.  
We allow the scattering extinction to act as a pure loss term for now.

Table~\ref{tab:AllFits} shows the fits to both the annular spectrum and the readout streak, when available.  We chose not to tie the $\NH$ column among all observations, to evaluate biases in $\NH$, spectral index, and the search for a black body component 
that could present itself in some observations.  
We also seek to compare our search for a thermal component of the spectrum to those of P16.  Several conclusions can be drawn from these results.

First, as seen in Figure~\ref{fig:SourceSpectrum}, the annular spectrum of \newflare\ has a soft X-ray excess in comparison to the readout streak spectrum.  This is due to contamination from the dust scattering halo, which is brighter at soft energies.  This leads to a systematically lower $\NH$ column measured from the annular spectrum, as compared to the readout streak spectra.  
A second effect is that the photon index on the annular fits (power law only) is systematically larger than in the readout streak spectra, and much higher than typical for X-ray binaries.  A steep spectral slope is what led P16 to conclude that there is a thermal component to the spectrum.  Regardless, both the annular and readout streak (power law only) spectra are consistent with the photon index measured by P16 for the \Swift\ XRT observations performed in the first few weeks after the \newflare\ outburst discovery.  In that paper, they conclude that \newflare\ was in a hard state at the time of \Chandra\ ObsIDs 18055 and 18056.  Yet these two observations also happen to occur at a time when the object's spectrum appeared to be softening dramatically, even though P16 conclude that it did not fully reach a steady soft state until ten days later (Figure~6 of P16).

Another interesting conclusion that can be drawn from Table~\ref{tab:AllFits} is that the readout streak spectrum is well described by either a pure power law or pure black body continuum spectrum.  The results of a power law plus black body are inconclusive, or at least not well constrained.  This ambiguity may be because there are significantly fewer counts in the readout streak ($\sim$5,000 counts) in comparison to the annular spectra ($\sim$18,000 counts).  Regardless, for the dust-contaminated annular spectra in ObsID 18055 and 18056, neither a single power law nor a single black body model fits well alone.  The combined model requires a low temperature component, $kT = 0.4$~keV.  We caution that for all the annular spectra, the apparent soft black body component may be an artifact of the dust scattering halo.

The $\NH$ values from the readout streak (power law only) is similar with the $\NH$ extinction measured for \SgrA \citep{Baganoff2003, Nowak2012}, supporting the assumption that \newflare\ is indeed a compact object in the GC.  If we take the single temperature black body spectrum, which could originate from a neutron star thermonuclear flare, then we get a smaller ISM column, $\NH \approx9 \times 10^{22}$~cm$^{-2}$.  
However, given the fact that thermonuclear flares are very short in duration ($\sim 10-100$~ks), this single black body scenario is highly unlikely if not impossible.

The later two \Chandra\ observations, ObsID 18731 and 18732, go beyond the time frame covered in P16.  One thing is immediately apparent: the source exhibits a much harder spectrum.  However, the $\NH$ column obtained in these observations is also very different from the previous ones.  
If we fix the $\NH$ value in ObsID 18731 and 18732 to match the annular spectra in the first two observations ($\NH = 13.0 \times 10^{22}$~cm$^{-2}$), we get a similar behavior to ObsID 18055 and 18056.  
The best fit model is a power law with moderate photon index ($\Gamma=2$-3) plus a low temperature black body ($kT=0.2$-0.3).  
Comparing this to the differences seen between the annular and readout streak fits, it appears that the spectral shape has not changed considerably between the two \Chandra\ epochs.  These fits are consistent with the picture drawn by P16, where \newflare\ is a LMXB that transitioned from a hard (covered by \Swift/XRT and the first two Chandra observations) to soft state (observed by \Swift\ and XMM-Newton).  It is typical for these types of objects to eventually fade back into a low luminosity hard state, as seen in the last two Chandra observations.

	\section{Conclusions}
	\label{sec:Conclusions}

We have shown that the residual surface brightness surrounding \newflare\ is caused by X-ray scattering from dust in the Galactic Center foreground.  
Integrating the surface brightness profile in Figure~\ref{fig:ScatteringHalo}, we find that the 1-6~keV scattering halo within $\alpha < 100''$ has a flux that is 14-21\% of the apparent flux of \newflare.  For the representative ISM column of $\NH = 10^{23}$~cm$^{-2}$ we expect a 1-6~keV halo to be on the order of 30\% of the point source flux.  
The missing flux might be explained by both absorption of the scattering halo by foreground clouds and the fact that we are only examining a small portion of the full scattering halo, which can extend as far as $10'$ in radius.  We are limited by both the field of view of 1/8 subarray mode and extended X-ray emission from the GC, which can outshine dust scattering.

Regardless, the inner portion of the scattering halo ($r < 10''$) contains 6-12\% times the flux of the point source.  Extend this to our understanding of \SgrA, with the fiducial Bondi radius of $3.7''$ \citep{Wang2013}.  The brightness from the inner scattering halo is comparable to the fraction of quiescent \SgrA\ flux that has been attributed to unresolved flares \citep[10\%,][]{Neilsen2013} and to point-like emission \citep[4\%,][]{Roberts2016}.  

If we use the MRN dust grain size distribution, then the inner scattering halo could potentially be explained by a small amount of dust within 250~pc of the GC, accounting for about $3-5\%$ of the total ISM column.  
This speaks to the remarkable ability of X-ray scattering to probe small populations of dust grains that are otherwise undetectable.  However, to fully understand the dust scattering halo presented here, we need to investigate the evolution of the dust scattering halo over time to eliminate degeneracies between the position of the dust clouds and the dust grain size distribution.  We reserve this analysis for a future paper.

\newflare\ produced one of the most luminous X-ray outbursts in the history of GC monitoring.  
After applying pileup mitigation techniques, the flux values and spectral index we measured from the first two \Chandra\ observations of \newflare\ are consistent with P16.  
The last two \Chandra\ observations took place later and have a lower spectral index than the last observations published in P16.  Our results thereby  support the hypothesis that the object is a LMXB that is back in a low-hard state, completing one full outburst cycle.  
We have also shown that the method of extracting an annular spectrum is subject to problematic biases from dust scattering halos.

\vspace{0.1in}
We thank the anonymous referee for their suggested changes, which helped clarify several points throughout this work.  
Support for this work was provided by NASA through \Chandra\ Award Number TM17910568 and the Einstein Postdoctoral Fellowship grant number PF6-170149 awarded by the Chandra X-ray Center, which is operated by the Smithsonian Astrophysical Observatory for NASA under contract NAS8-03060.  The Guaranteed Time Observations (GTO) included here were selected by the ACIS Instrument Principal Investigator, Gordon P. Garmire, of the Huntingdon Institute for X-ray Astronomy, LLC, which is under contract to the Smithsonian Astrophysical Observatory; Contract SV2-82024.  B.M. and D.H. are supported by a Natural Sciences and Engineering Research Council of Canada Discovery Grant. B.M. is also supported by McGill's Rubin Gruber Science Undergraduate Research Award. D.H. is also supported by a Fonds de recherche du Québec - Nature et Technologies Nouveaux Chercheurs Grant.  F.K.B is supported by SAO contract SV2-82023 under NASA contract NAS8-03060.  N.D. is supported by a Vidi grant from the Netherlands Organization for Scientific research (NWO) and a Marie Curie fellowship (contract no. FP-PEOPLE-2013-IEF-627148) from the European Commission.

\bibliography{ext_letter,sgra,references}

\end{document}